# Collimated blue light generated by four-wave mixing in Rb vapour

Alexander M. Akulshin, Russell J. McLean, Andrei I. Sidorov, and Peter Hannaford

*Centre for Atom Optics and Ultrafast Spectroscopy, Swinburne University of Technology, Melbourne, Australia*

**Abstract**

We investigate frequency up-conversion of low power cw resonant radiation in Rb vapour as a function of various experimental parameters. We present evidence that the process of four wave mixing is responsible for unidirectional blue light generation and that the phase matching conditions along a light-induced waveguide determine the direction and divergence of the blue light. Velocity-selective excitation to the 5D level via step-wise and two-photon processes results in a Doppler-free dependence on the frequency detuning of the applied laser fields from the respective dipole-allowed transitions. Possible schemes for ultraviolet generation are discussed.

## 1. Introduction

The quantum mechanical effect of lasing without inversion (LWI) [1,2] has attracted considerable attention because of its potential in realizing sources of coherent radiation in the deep ultraviolet. To date, however, the frequency of new optical fields generated using the LWI approach exceeds the frequency of the pump radiation by only a few percent. Several different but related methods for frequency up-conversion in atomic media have been demonstrated. A blue laser based on alkali vapours with an almost doubling of the infrared pump frequency was proposed in [3]. The same degree of frequency up-conversion can be achieved using an alternative approach based on light-induced atomic coherence. Indeed, ground-state atomic coherence may facilitate the generation of new optical fields in the same spectral region as the applied radiation [4-6], while coherence prepared by near-infrared resonant radiation in an atomic medium with ladder-type energy levels can produce coherent blue light [7-9]. The blue light generation was attributed to atomic coherence in the cascade system; however, the specific mechanism of the effect was not discussed in detail. It was also emphasized in [7] that this effect is different to two-photon parametric gain. Recently, blue light obtained in a cascade system was used for state-selective imaging of cold atoms [10].

Our experimental investigations of this relatively unexploited scheme of frequency up-conversion are primarily motivated by possible applications in quantum-information science, although extensions of the approach to single-atom detection and ultraviolet generation appear promising. This paper is aimed at understanding the primary processes responsible for the blue light generation by studying its spatial and spectral characteristics at relatively low cw laser power and atomic density.

## 2. Excitation scheme and experimental setup

We study frequency up-conversion of near-IR cw resonant laser radiation in a warm rubidium vapour cell containing a natural mixture of $^{85}$Rb and $^{87}$Rb isotopes. The relevant optical transitions are shown in Fig. 1a. Rubidium atoms, after excitation to the $5D_{5/2}$ state by absorbing photons at 780 nm and 776 nm, decay with a 35% probability to the intermediate $6P_{3/2}$ state and then with a 31% probability to the ground state [11], emitting blue photons at 420 nm, as demonstrated in Fig. 1b.

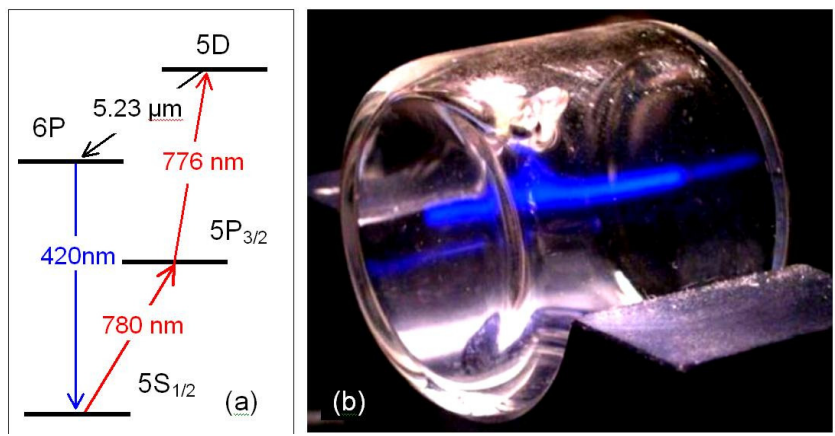

Fig. 1. (a) Energy levels of Rb atoms and the optical transitions used in the experiments. (b) Blue fluorescence of Rb atoms excited by the bichromatic resonant light.

Both counter- and co-propagating geometries for the applied laser beams have been used for studying the *5D* level excitation, which was controlled by monitoring the resonant fluorescence at 420 nm. Despite the fact that the blue fluorescence is much stronger for the counter-propagating case [12] collimated blue light (CBL) appears only for the co-propagating geometry and at certain atomic densities, intensities and detunings of the applied bichromatic laser light. Peculiarities of Doppler-free two-photon spectroscopy with bichromatic nearly resonant light will be discussed elsewhere. It was found in [6] that the CBL generation takes place only if the far-IR radiation corresponding to the $5D \rightarrow 6P$ transition is present. During our experiment a detector for IR radiation was not available and we were able to analyze only the blue light.

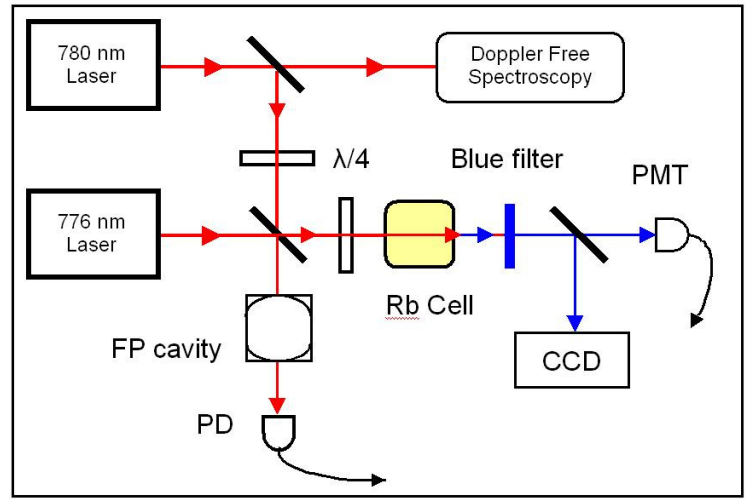

Fig. 2. Scheme of the experiment.

A simplified scheme of the experimental setup is shown in Fig. 2. Two diode laser systems have output powers of 20 mW and 15 mW at 780 nm and 776 nm, respectively. Both lasers have short-term linewidths of approximately 1 MHz. Collimated radiation from both lasers is combined in a bichromatic beam that propagates along the axial direction of the 5 cm-long heated cells. The cross-section and intensity of both components of the bichromatic beam are controlled with lenses and neutral density filters. The cross sections of the laser beams are measured by the knife-edge method. We also use a single-mode optical fibre to achieve perfect overlapping of the components of the bichromatic beam.

Colour and interference filters with high transmission at 420 nm and optical densities in the range 1.0 to 3.0 at 780 nm and 776 nm are used to select the collimated blue light and the isotropic resonant fluorescence of Rb atoms at 420 nm, which are detected by Hamamatsu photomultiplier tubes (PMT) and recorded using a digital oscilloscope. The spectra are obtained by scanning the optical frequency of one laser while keeping the frequency of the other laser fixed. The frequency of the 780 nm laser is scanned across the $D_2$ absorption line or locked to Doppler-free resonances produced in an auxiliary

Rb cell, while the frequency of the 776 nm laser is controlled with a tunable low-finesse Fabry-Perot cavity or scanned across the $5P_{3/2} \rightarrow 5D$ transitions. Spectra of the collimated blue light are also analyzed by an AQ-6315E Anritsu optical spectrum analyzer with 0.05 nm resolution.

### 3. Results and discussions

*3.1 Four-wave mixing and spatial characteristics*

We explain the CBL generation as a result of four-wave mixing in a diamond-type energy level configuration. Usually for non-degenerate four-wave mixing in atomic media three laser fields are applied to produce a new optical field [12]. In the present scheme only two laser fields, at 780 nm and 776 nm, are applied and the third field is produced by amplified spontaneous emission resulting from population inversion on the 5D $\rightarrow$ 6P transition along paths containing 5D atoms. The estimated wavelength of the far-IR radiation, which corresponds to the energy difference between the 5D and 6P levels, is 5.23 μm. This radiation is expected to be very anisotropic with maxima in both the co- and counter-propagating directions because of the elongated shape of the atom-light interaction region defined by the applied laser beams and where population inversion occurs. The three optical fields consequently produce radiation at 420 nm by four-wave mixing in a direction which satisfies the phase-matching relation $k_1 + k_2 = k_{IR} + k_{BL}$, where $k_1$, $k_2$, $k_{IR}$ and $k_{BL}$ are the wave vectors of the radiation at 780 nm, 776 nm, 5.23 μm and 420 nm, respectively. This relation can be satisfied only for co-propagating far-infrared radiation. Unidirectional generation of the CBL is confirmed experimentally. This is strong evidence for the four-wave mixing mechanism of CBL generation.

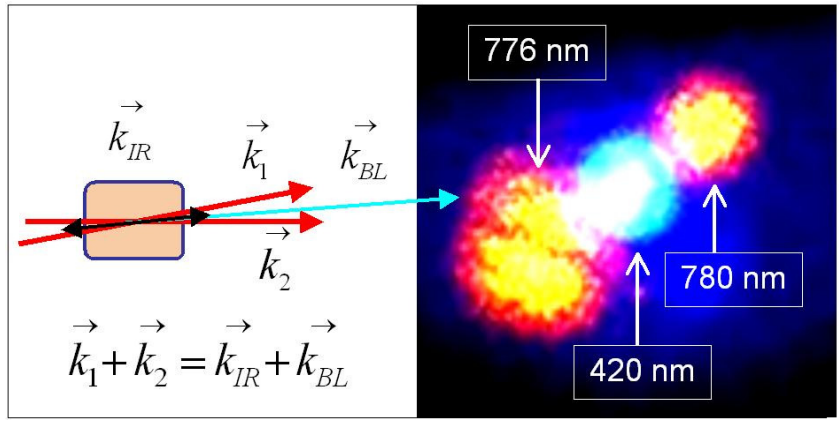

Fig. 3. Scheme which illustrates the phase-matching relation for nearly-parallel laser beams and observed profiles of the light beams after the cell

superimposed on isotropic fluorescence background recorded using a spatially sensitive detector.

If the laser beams overlap in the cell at a small angle, then the laser beams and the CBL are spatially separated beyond the cell. We find that the direction of the blue light coincides with the direction expected from the phase-matching relation (Fig. 3). This is a further indication of the four-wave mixing origin of CBL generation; although, the possibility of a competing parametric process that results in simultaneous emission of far-infrared and blue photons cannot be excluded.

When the applied laser beams are perfectly overlapped by coupling to a single mode fibre, the CBL is generated exactly in the forward direction. The minimum measured divergence of CBL is less than 5 mrad. We also found that the divergence and even the transverse mode pattern of the blue light is sensitive to the optical frequency of the applied laser light, as shown in Fig. 4. The observed profiles of the CBL have a similar appearance to laser or optical fibre transverse modes. We attribute these patterns to the properties of a light-induced waveguide formed in the vapour by refractive index variations seen by the optical fields, but the transverse structure of the CBL needs further investigation.

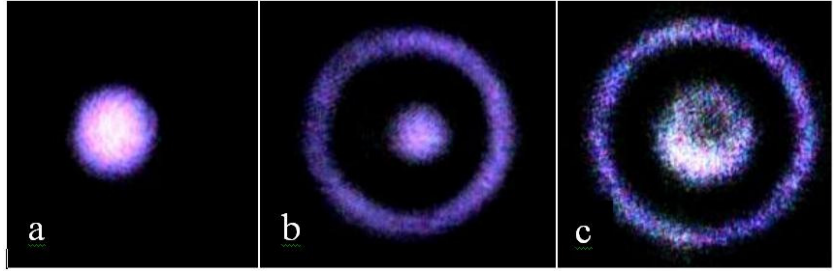

Fig. 4. Cross sections of the blue light beam observed at different frequencies of the 780-nm laser in the vicinity of the $5S_{1/2}$ ($F=3$)$\to$ $5P_{3/2}$ transitions in $^{85}$Rb. (a), (b) and (c) correspond to the laser frequency tuned to the $5S_{1/2}$ ($F=3$) $\to$ $5P_{3/2}$ ($F'=2$) transition, $5S_{1/2}$ ($F=3$) $\to$ $5P_{3/2}$ ($F'=4$) transition and to high-frequency slope of the Doppler profile, respectively. The 776-nm laser frequency is tuned to a maximum of the CBL.

*3.2 Frequency detuning*

We have investigated the influence on the CBL power of the frequency detuning of the components of the bichromatic beam from the corresponding atomic transitions. The efficiency of the blue light generation depends on the population of the $5D$ level. The transfer of atoms from the ground state to the $5D$ level can occur via step-wise or two-photon processes, i.e., by the absorption of photons from the resonant fields sequentially via the two dipole-allowed transitions ($5S_{1/2}\to5P_{3/2}\to5D_{5/2}$) or simultaneously through the two-photon transition $5S_{1/2} \rightarrow 5D_{5/2}$. The excitation to the *D* level by laser fields which are nearly resonant to the intermediate level has been analyzed both theoretically and experimentally, for example, in [12]. Step-wise excitation occurs if one laser field is tuned to resonance with the intermediate level, while for two-photon excitation the sum ($v_1+v_2$) corresponds to the frequency of the $5S_{1/2}(F=3) \rightarrow 5D_{5/2}$ transition in [85]Rb. The excitation rates for both processes have different amplitudes and widths; however, the excitation is greatly enhanced when both laser fields are in exact resonance with the intermediate level.

If the 780-nm component is tuned to the Doppler broadened $5S_{1/2}(F=3) \rightarrow 5P_{3/2}(F'=2, 3, 4)$ transitions in [85]Rb, the radiation is in exact resonance simultaneously with atoms within three velocity groups (ignoring the hyperfine structure of the $5D_{5/2}$ level). However, two of the groups are depopulated by hyperfine optical pumping, hence only in the third group, in which the interaction occurs on the cycling transition $5S_{1/2}(F=3) \rightarrow 5P_{3/2}(F'=4)$, the excitation to the *D* level is efficient. This occurs when the sum ($v_1+v_2$) corresponds to the frequency of the $5S_{1/2}(F=3) \rightarrow 5D_{5/2}$ transition. Detailed considerations of velocity selective optical pumping produced and probed with two independent lasers in *N*- and ladder-type atoms can be found in [3, 11]. The velocity-selective excitation to the $5D_{5/2}$ level is demonstrated in Fig. 5.

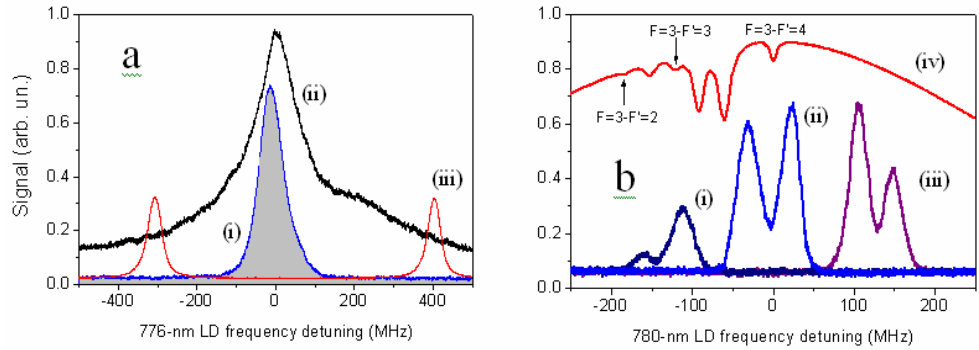

Fig. 5. (a) Collimated blue light (i) and fluorescence at 420 nm (ii) as a function of the 776-nm laser detuning observed at 60 [0]C. The 780-nm laser is locked to the $5S_{1/2}(F=3) \rightarrow 5P_{3/2}(F'=4)$ transition in [85]Rb, while the 776-nm laser is scanned in the vicinity of the $5P_{3/2}(F'=4) \rightarrow 5D_{5/2}$ transition. Curve (iii) shows transmission resonances of the Fabry-Perot cavity. (b) Collimated blue light generated in the Rb cell at 82 [0]C vs the 780-nm laser frequency detuning from the $5S_{1/2}(F=3) \rightarrow 5P_{3/2}(F'=4)$ transition. Curves (i), (ii) and (iii) correspond to different fixed frequencies of the 776 nm laser. Curve (iv) represents the reference Doppler-free absorption resonances observed in an auxiliary Rb cell.

High laser power is not needed for CBL generation if the bichromatic beam is focused in the Rb vapor cell. Figure 5a shows spectral profiles of both CBL and blue fluorescence obtained in the Rb cell when the minimum diameter of the combined bichromatic beam, consisting of 0.17 mW at 780 nm and 0.65 mW at 776 nm, is approximately 120 µm. The frequency of the 780-nm laser is tuned and locked to a Doppler-free resonance on the $5S_{1/2}(F=3) \to 5P_{3/2}(F'=4)$ transition in $^{85}$Rb, while the 776-nm laser is scanned across the $5P_{3/2} \to 5D_{5/2}$ transition. Both CBL and blue fluorescence have a Doppler-free dependence on the frequency detuning of the applied laser fields from the corresponding transitions. However, the shapes of the resonance are different and the maximum of the CBL profile is shifted from the fluorescence maximum.

At higher light intensity the CBL spectra reveal a doublet structure. Figure 5b shows several profiles of CBL as a function of the 780-nm laser detuning from the $5S_{1/2}(F=3) \to 5P_{3/2}(F'=4)$ transition in $^{85}$Rb. Now the frequency of the 776-nm laser is fixed in the vicinity of the $5P_{3/2}(F'=4) \to 5D_{5/2}$ transition. The intensity-dependent splitting suggests that the doublet can be attributed to the Autler-Townes effect. The 776 nm light couples the $5P_{3/2}(F'=4)$ and $5D_{5/2}$ levels producing ac Stark splitting in both of them. CBL peaks occur when the 780 nm laser light is resonant with the transition from the $5S_{1/2}(F=3)$ ground state to either ac-Stark split component of the $5P_{3/2}(F'=4)$ level. Figure 5b also shows that the maximum CBL power occurs when the 780 nm laser light is tuned to the cycling transition, confirming that it makes the major contribution to the frequency up-conversion process.

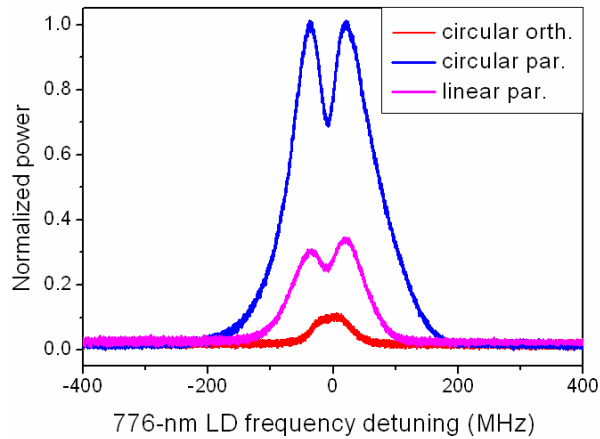

Fig. 6. Profiles of the CBL at different polarizations as a function of the 776-nm laser frequency in the vicinity of the $5P_{3/2} \to 5D_{5/2}$ transitions in $^{85}$Rb. The 780-nm laser is locked to the $5S_{1/2}(F=3) \to 5P_{3/2}(F'=4)$ transition.

*3.3 Polarization dependence*

Both the maximum power and line shape of the CBL are sensitive to the polarizations of the applied laser fields. The up-conversion process is sensitive to the distribution of atoms among the hyperfine and Zeeman sub-levels of the 5$D$ level. We find that more efficient CBL generation occurs if the excited atoms are concentrated in a few specific Zeeman sub-levels. This occurs when both components of the bichromatic radiation have the same circular polarization, which tends to pump atoms into the state with maximum |$m_F$|. This also means that the excitation pathway uses transitions with the highest probability. Decay from the 5$D_{5/2}$ ($F''$=5, $m$=+5) level is possible only to the 6$P_{3/2}$($F'$=4, $m$=+4) and then to the 5$S_{1/2}$ ($F$=3, $m$=+3) levels. The CBL power is almost ten times higher in the case of the same circular polarizations of the applied fields, as shown in Fig. 6.

Angular momentum conservation during the wave-mixing process explains the observed predominantly circular polarization of CBL in the case of same circular polarizations of the applied laser fields.

*3.4 Intensity dependence*

The CBL power depends on the power of the applied laser radiation, as shown in Fig. 7. The data are taken at relatively low atomic density. The bichromatic beam, which consists of circularly polarized components of equal power, is focused to a spot of diameter 120 μm, so that the maximum intensity of each component was approximately 4.7 W/cm$^2$. Both dependences presented in Fig. 7a exhibit a nonlinear behavior and increase faster than the steady-state excitation rate of ladder-type atoms at rest in the low-intensity limit [12]. Plotting the experimental data on logarithmic scales suggests that the CBL power growths can be approximated by a 1.5 power dependence on the applied laser power.

The dependence of the CBL power on the power at each resonant wavelength (with the power of the other component fixed) was also analyzed. Both power dependences shown in Fig. 7b reveal clear thresholds, which are approximately 800 mW/cm$^2$ and 70 mW/cm$^2$ for the 780-nm and 776-nm components, respectively. Also, the CBL dependence on the 780-nm component is much steeper, especially below 0.12 mW, where the CBL power scales as input power to the power of approximately 6.5. The difference is explained by the different absorption at the two wavelengths in the cell. Significant absorption of the 776-nm component occurs in the case of efficient population transfer to the 5$P_{3/2}$ level along the entire length of the atomic sample, while for the 780-nm component a linear absorption length at this temperature is less than the length of the cell. Thus, light intensity at 780 nm well above the saturation intensity for the 5$S_{1/2}$($F$=3)→5$P_{3/2}$($F'$=4) transition is needed to make the atomic sample transparent enough to the 780-nm light. A higher intensity of the 780-nm component increases both the interaction depth and the excitation rate, making the CBL power dependence very steep.

The maximum power of blue light obtained in the vapor cell at 87 $^0$C with circular polarization of the applied bichromatic light but without thorough optimization was 15 µW.

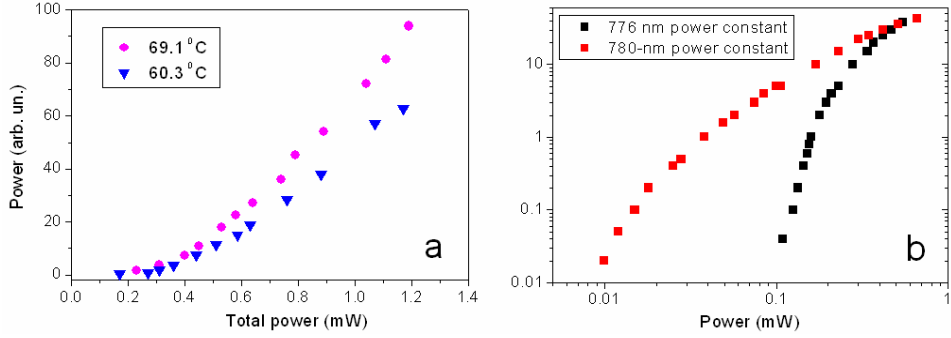

Fig. 7. Power dependences of the CBL obtained with circularly polarized bichromatic light when the frequency of the 780-nm laser is locked to the $5S_{1/2}$ ($F$=3)→$5P_{3/2}$($F'$=4) transition, while the 776-nm laser is tuned to the maximum power of the blue light. (a) Power of the CBL at different atomic densities as a function of the total applied laser power. Both components of the bichromatic beam have equal power. (b) CBL power at an atomic density of 3 × 10$^{11}$ cm$^{-3}$ (60.3 $^0$C) as a function of the power of each component, while the other component is kept unchanged at 0.65 mW.

*3.5 Atomic density dependence*

The atomic density strongly affects the power and spectral properties of the CBL. The temperature dependence of the blue light power is shown in Fig. 8. The number density $N$ is estimated using the Killian formula [14] and the temperature of the coldest part of the cell. The minimum atomic density at which CBL is detected is approximately 1.7 × 10$^{11}$ cm$^{-3}$. The blue light power grows sharply in the temperature range from 53 $^0$C to 58 $^0$C, and then slows down, until above 70 $^0$C ($N \approx 8\times10^{11}$ cm$^{-3}$) the CBL power is saturated. The values of both saturation and threshold atomic densities depend on the applied light intensity. At higher laser intensity CBL can be obtained at lower atomic density and CBL saturation occurs at higher atomic density.

Because of the high absorption of the 780-nm resonant light at atomic densities above $N \simeq 1\times10^{12}$ cm$^{-3}$, the efficiency of the up-conversion process becomes very inhomogeneous along the cell and Rb atoms at the far end of the cell predominantly absorb CBL rather than contributing to its generation.

At relatively low atomic density ($N \leq 1\times10^{12}$ cm$^{-3}$), blue light can be obtained only in the vicinity of the strongest cycling transition within the $^{85}$Rb $D_2$ line, while at higher atomic density CBL occurs when the same laser is tuned to either Rb isotope, as shown in Fig. 9, curve (i). The CBL generation at two different frequencies of the 780-nm laser and a single frequency of the 776-nm laser is possible because of the small isotope shift of the $5D_{5/2}$ level. The width of the spectral profile of the blue light increases with atomic density from approximately 65 MHz to approximately 500 MHz.

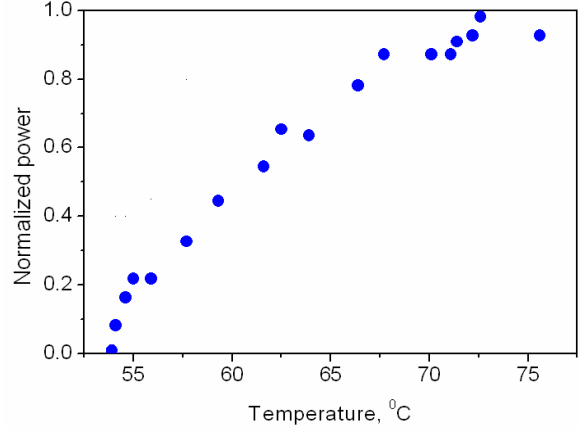

Fig. 8. Temperature dependence of the CBL. The 780-nm laser is locked to the $5S_{1/2}$ ($F=3$) → $5P_{3/2}$ ($F'=4$) transition in $^{85}$Rb while the 776-nm laser is tuned to a maximum of the CBL.

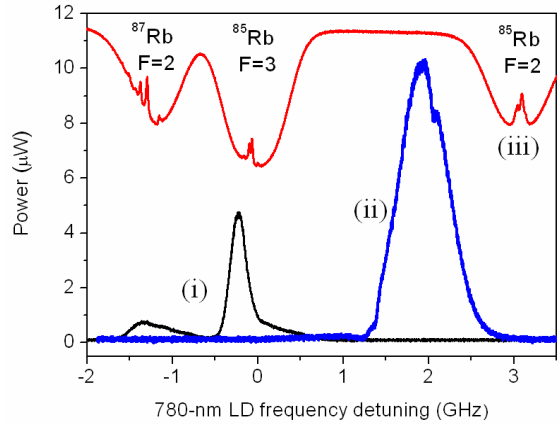

Fig. 9. Profiles of CBL generated at different atomic densities vs the 780-nm laser frequency detuning from the $5S_{1/2}$ ($F=3$) → $5P_{3/2}$ ($F'=4$) transition $^{85}$Rb, while the frequency of the 776-nm laser is tuned to the maximum of the blue light power. Curves (i) (multiplied by 5) and (ii) were obtained at $N = 0.3 \times 10^{12}$ cm$^{-3}$ and $N = 1.4 \times 10^{12}$ cm$^{-3}$, respectively, while curve (iii) shows the reference absorption profile in the auxiliary Rb cell.

At atomic densities above $N = 1.3 \times 10^{12}$ cm$^{-3}$ CBL cannot be generated when the 780-nm laser is tuned in the vicinity of the $5S_{1/2}(F=3) \rightarrow 5P_{3/2}$ transitions of $^{85}$Rb. Instead, the maximum power of the blue light occurs when the 780-nm laser is tuned to the spectral region between the two absorption lines of $^{85}$Rb, corresponding to the $5S_{1/2}(F = 2) \rightarrow 5P_{3/2}$ and $5S_{1/2}(F=3) \rightarrow 5P_{3/2}$ transitions (Fig. 8, curve (ii)). A similar shift was observed at a temperature of above 140 $^0$C in [8], where, using a semiclassical approach based on the Bloch equations and a five-level model, it was shown that the gain at 420 nm has a maximum when the 780-nm laser is tuned between the two Doppler-broadened absorption lines.

Another possible explanation of this shift is that it represents a transition from the regime where the stepwise and two-photon excitation processes are both occurring and are indistinguishable, to the regime where the two-photon excitation mechanism is dominant. As the resonant absorption of 780-nm light increases, its penetration into the cell decreases so that blue light is generated by the stepwise process only near the front end of the cell and absorbed before reaching the other end, while two-photon excitation to the $5D$ level can occur for non-resonant 780-nm light over the entire length of the cell even at high atomic density.

## 3.6 Fine structure of the 5D level

It is worth noting that tuning the 776-nm laser to reach the $5D_{3/2}$ level instead of $5D_{5/2}$, which are separated by 0.18 nm, results in a reduction by about an order of magnitude of the CBL power; however, the wavelength of the CBL, which corresponds to the $6P_{3/2} \rightarrow 5S_{1/2}$ transition, remains unchanged. Atoms from the $5D_{3/2}$ level can decay via either fine structure component of the $6P$ level, and then to the ground state by the dipole-allowed transitions $6P_{1/2} \rightarrow 5S_{1/2}$ and $6P_{3/2} \rightarrow 5S_{1/2}$ with a wavelength separation of approximately 1.4 nm. However, the CBL spectrum reveals no fine structure, suggesting that the threshold conditions for CBL generation via the $6P_{1/2}$ path are not met. This may be important for quantitative modeling of the process of blue light generation.

The observation of resonant fluorescence excited by CBL alone in an additional Rb cell shows that the optical frequency of the blue light is detuned from the $5S_{1/2} \rightarrow 6P_{3/2}$ transition by less than the Doppler width $\Delta \nu_D \simeq 0.9$ GHz for this transition.

## 3.7 Possible schemes for UV generation

Ultraviolet generation using this up-conversion technique in an alkali-metal vapour should be possible. The recent demonstration of coherent blue light for the equivalent level scheme in cesium [9], where the branching ratio is very unfavorable (only 0.14% of atoms excited to the uppermost $6D_{5/2}$ level decay to the ground state emitting blue photons), suggests that ultraviolet generation in an alkali-metal vapour should be possible. Indeed, excitation to the Rb $6D_{5/2}$ level by resonant bichromatic radiation at 780 nm and 630 nm may result in a new field generation at 357.8 nm on the $7P_{3/2} \rightarrow 5S_{1/2}$ transition, where the decay path has a probability higher than 0.14%. Coherent frequency tunable radiation at these wavelengths is readily obtained from diode lasers. Even higher energy levels of Rb atoms could be reached via step-wise and two-photon processes using 420 nm light as the first excitation step.

## 4. Conclusion

The characteristics of frequency up-conversion of low power cw laser radiation by the process of resonant wave mixing in a Rb vapour cell at relatively low atomic density ($1.3 \times 10^{11}$ cm$^{-3}$ ≤ N ≤ $2 \times 10^{12}$ cm$^{-3}$) have been investigated. We find that, due to velocity-selective excitation to the $5D$ level, the collimated blue light has a Doppler-free dependence on the frequency detuning of the applied laser fields from the corresponding dipole-allowed transitions under conditions of perfect overlapping of the laser beams and small power broadening. Unlike that for blue resonant fluorescence, the process of blue light generation has a threshold-like behavior with respect to the atomic density and the applied light intensity. The spatial and spectral properties of the blue light are very sensitive to parameters that include the frequency detuning of both lasers, the polarizations of the applied laser fields and their spatial overlap.

We explain the CBL generation as a result of four-wave mixing in a diamond-type energy level configuration. The phase matching determined by the wave vectors of all the optical fields involved in the wave-mixing process and the refractive index seen by all the fields is a crucial parameter for the CBL generation. The many parameters involved and the complex nonlinear properties of atomic media make the process of blue light power optimization difficult but interesting. The transition from the indistinguishable stepwise and two-photon excitation process to the off-resonant two-photon dominant excitation mechanism requires an additional investigation. Possible schemes for ultraviolet generation using this approach are discussed. Further research promises new interesting and potentially useful features of this technique.

## 5. Acknowledgements

We thank Dr. Mandip Singh for his contributions at the early stages of the experiment and Russell Andersen for useful discussions.